# Slowing evolution is more effective than enhancing drug development for managing resistance

Nathan S. McClure[1] and Troy Day[1,2]

1. Department of Biology, Queen's University, Kingston, ON, K7L 3N6, Canada
2. Department of Mathematics and Statistics, Queen's University, Kingston, ON, K7L 3N6, Canada

**Drug resistance is a serious public health problem that threatens to thwart our ability to treat many infectious diseases [1-4]. Repeatedly, the introduction of new drugs has been followed by the evolution of resistance [5-12]. In principle there are two ways to address this problem – (i) enhancing drug development, and (ii) slowing drug resistance. We present data and a modeling approach based on queueing theory that explores how interventions aimed at these two facets affect the ability of the entire drug supply system to provide service. Analytical and simulation-based results show that, all else equal, slowing the evolution of drug resistance is more effective at ensuring an adequate supply of effective drugs than is enhancing the rate at which new drugs are developed. This lends support to the idea that evolution management is not only a significant component of the solution to the problem of drug resistance, but may in fact be the most important component.**

In principle, the evolution of resistance to any particular drug is not problematic provided that an alternative drug is available. What matters is therefore the rate at which drug resistance evolves relative to the rate at which new drugs are brought to market. Consequently there are two ways to ensure the availability of effective drugs: (i) by increasing the rate of drug discovery, or (ii) by increasing the time it takes for resistance to evolve through better resistance management.

Approaches for increasing the rate of drug discovery are probably familiar, and include research devoted to developing screening technology for new compounds as well as developing new classes of antimicrobial agents. Approaches for slowing the evolution of resistance are perhaps less familiar, but include research into new strategies for reducing the inappropriate use of antimicrobials, determining when and where existing drugs should be used in combination versus as sequential monotherapies, as well as determining the optimal dose and timing of deployment for these existing drugs [1, 13-16]. Although precise estimates are difficult to come by, it would appear that the research effort devoted to drug discovery currently far exceeds that devoted to resistance management [1].

Is this current disparity in the effort devoted to drug discovery versus resistance management an effective use of resources? To answer this question we need to determine the benefits, in terms of ensuring effective drug availability, of increasing the rate of drug discovery versus slowing the rate of evolution through better resistance management. At one level the answer to this question is obvious. If there is an upper limit to the number of drugs that can be developed for a particular disease, then at some point drug development will become effectively impossible. This would leave slowing evolution as the only option. But what if we are not yet facing this limitation on drug development? Here we show that even when there is no limit on the development of new drugs, slowing evolution is still inherently more effective. This thereby calls into question the current emphasis on the drug discovery side of the issue, and suggests that more emphasis ought to be given to evolution management.

Historical data for the development of new antimalarial and antibiotic drugs, as well as the evolution of resistance to these, is presented in Figure 1. The processes underlying these data are complex, with both geographical and temporal variation in the drugs that are used to treat specific infectious diseases. We construct and analyze a simplified model of these processes that abstracts only the essential features. The results we present are generic (i.e., not specific to any particular drugs or diseases) but we illustrate them with specific examples whose parameter values are taken from data on malaria [6-12].

We define a "drug" as any whole treatment strategy, which might consist of one or multiple active ingredients. The "drug portfolio" is the collection of existing drugs that are effective against a specific type of infection. For simplicity we assume that a single drug is used at any given time, and its use is continued until resistance to this drug has appeared and reached some threshold frequency (e.g., 10% see [13]). The time it takes for this to occur is referred to as the 'time to evolve resistance', or equivalently the 'drug lifespan'. It is denoted by the random variable $L$. At this point the use of the drug is effectively discontinued and another drug from the drug portfolio, if available, is then brought into use. During this process, new drugs are also in various stages of development and are occasionally brought into the drug portfolio. Although this is clearly a simplified description of reality, relaxing some of these assumptions (including allowing for multiple drugs to be used simultaneously) does not alter the important qualitative insights provided by our analysis (see Appendix).

The overall dynamics of the modeled drug supply system are shown schematically in Figure 2. Drug arrival into the drug portfolio, and the evolution of resistance both occur stochastically, resulting in periods of time for which effective treatment is available (i.e., when there is at least one effective drug in the drug portfolio), separated by periods when there is no effective treatment. We refer to the length of time during which effective treatment is available as the "time to failure" (denoted by $T$). These periods are separated by time intervals during which no effective treatment is available (Figure 2). For example, the appearance of pan-resistant *Klebsiella pneumoniae* and multidrug resistant Tuberculosis suggest that we might well have exceeded the time to failure for these diseases, and be heading into a period in which no effective treatment is immediately available.

We can view the expected time to failure as the product of the expected number of drugs used before failure occurs, and the expected lifespan of each of these drugs. We will also be interested in the long-run fraction of time that effective treatment is available, and refer to this as the "drug availability". The model is completely specified by the process of drug arrivals and the process of resistance evolution.

Recent analysis of a large data set for pharmaceutical production from companies during the period 1950-2008 has shown that the annual output of new drugs is Poisson distributed with a constant rate parameter [17]. This implies that the time between new drug arrivals is exponentially distributed. Consequently we define the random variable $D$ to be the time between drug arrivals into the drug portfolio, and assume that $D$ is exponentially distributed with rate parameter, $\alpha$. The expected time between drug arrivals is therefore $E[D] = 1/\alpha$.

The time to evolve resistance (i.e., the drug lifespan, $L$) represents the time between when a drug is first used and when resistance to the drug reaches a threshold frequency. We assume that this lifespan is determined by the evolution of resistance, and therefore different evolution management strategies will result in different drug lifespans. Little is currently known about the distribution of $L$ but we can make some progress by examining the data from Figure 1. Analyses suggest that, for some diseases at least (e.g., malaria), $L$ is also approximately exponentially distributed (see Appendix). There are many caveats associated with this conclusion, however, and therefore we consider two scenarios. First, we suppose that *L* is exponentially distributed with rate parameter $\beta$. The expected drug lifespan is therefore $E[L] = 1/\beta$. Second, we consider a case where the distribution of *L* is left arbitrary.

We begin the analysis by assuming that the average time between drug arrivals is larger than the average drug lifespan (i.e., $E[D] > E[L]$). This assumption is relaxed when we consider a variable rate of drug development. This implies that, with probability 1, there will be periods of time when drugs are available as well as periods of time when they are not (Figure 2).

Two interventions are explored: (i) increasing the expected drug lifespan $E[L]$ through better evolution management, or (ii) decreasing the expected time between drug arrivals $E[D]$ through enhanced drug discovery. In the additive case we consider increasing $E[L]$ by an additive amount or decreasing $E[D]$ by the same amount. In the multiplicative case we consider increasing $E[L]$ by a *factor* or decreasing $E[D]$ by the same factor. For completeness we explore both additive and multiplicative changes in each. We focus on two main system performance measures: the time to failure, *T*, which is a random variable, and the drug availability, $\rho$ (Figure 2).

When the distribution of the time to evolution, *L*, is exponential an explicit equation can be derived for the probability density of the time to failure, *T*. We obtain

$$f_T(t) = \sqrt{\frac{\beta}{\alpha}} \frac{e^{-(\alpha+\beta)t} I_1(2\sqrt{\alpha\beta}t)}{t} \tag{1}$$

where $I_1(x)$ is a modified Bessel function of the first kind (see Appendix). Likewise, drug availability is given by

$$\rho = \frac{\alpha}{\beta}. \tag{2}$$

Increasing the mean time to evolve resistance (i.e., increasing $E[L] = 1/\beta$) or decreasing the mean time between drug arrivals (i.e., decreasing $E[D] = 1/\alpha$) both shift probability mass in equation (1) from low to high values of *T*. However, changes in the mean time to evolve resistance do so to a greater extent (Figure 3a). This is true regardless of whether the changes are additive or multiplicative (see Appendix). Likewise, drug availability, $\rho$, is also increased more through an additive change in the mean time to evolve resistance, whereas both interventions have identical effects on $\rho$ when the changes are multiplicative (see Appendix).

When the distribution of time to evolve resistance is left arbitrary we can obtain an equation for an integral transform of the distribution of time to failure, from which we can calculate any of its moments (see Appendix). We focus here on the first moment (i.e., expected time to failure), which is

$$E[T] = \frac{E[L]}{1 - E[L]/E[D]}. \tag{3}$$

The expression for drug availability in this case is identical to equation (2) (see Appendix). Again we can see that the expected time to failure, $E[T]$, increases more with a change in $E[L]$ than it does with a change in $E[D]$ (Figure 3b). And again this is true regardless of whether the changes are additive or multiplicative. The conclusions about drug availability are identical to the case where the distribution of time to evolution is exponential.

We also explored the case where the rate of drug development $\alpha$ varies as an inverse function of the drug portfolio size (see Appendix). We studied this situation to better understand the consequences of having drug development respond to drug demand. It also allows us to examine a case where the mean time between drug arrivals is shorter than the mean drug lifespan (i.e., $E[D]<E[L]$). Again, the results show that slowing evolution increases the expected time to failure and drug availability more than decreasing the time between drug arrivals (see Appendix).

**Discussion:**

Our results illustrate that slowing the evolution of drug resistance has a greater effect on the performance of the drug supply system than does speeding up drug development. What is the underlying mechanism behind this result?

Both types of interventions increase the chance of developing new drugs before the system as a whole fails. Decreasing the time between drug arrivals does so because, on average, more drugs will arrive in a given period of time. Increasing the time to evolve resistance does so because it extends the window of opportunity for a new drug to be brought to market before failure occurs. In fact, when the changes for each intervention are multiplicative it can be shown that, on average, the same number of drugs will arrive and be used before failure occurs in both cases. The difference lies in how the interventions affect the lifespan of each drug. Recall that we can express the mean time to failure, $E[T]$, as the product of the expected number of drugs that are used before failure and the expected lifespan of each drug. Mathematically,

$$E[T] = E[N] \times E[L] \qquad (4)$$

where $N$ is the number of drug used before system failure. As already described, multiplicative changes in both interventions have identical effects on $E[N]$. But changes in the time to evolve resistance also affect $E[L]$ whereas changes in drug development do not. As a result, increasing the expected time to evolve resistance has a compounding effect that is absent when changing the drug development time. Moreover, this effect is even greater when we consider additive changes in each intervention because changing the time to evolve resistance increases $E[N]$ more than does changing the time for drug development.

Drug availability is the proportion of time for which there is effective treatment. Hence, even though there is a greater increase in time to failure when evolution is slowed, it is important to consider how enhancing the rate of drug arrivals reduces the fraction of time for which there is no effective treatment. In fact, as a result of a multiplicative change, the increase in time to failure from slowing evolution of resistance is equivalent to the increase in drug availability from speeding up drug development. However, if the change is additive, the added benefit to time to failure from slowing evolution exceeds the effect that decreasing time between drug arrivals has on drug availability, causing drug availability to increase more when evolution of resistance is slowed.

Our simplified model of the drug supply system assumes that, when available, drugs are used one at a time. However, the data in Figure 1 suggests that multiple drugs are often used simultaneously at the population level (although perhaps in different geographic regions [18, 19]]. Modeling the simultaneous use of multiple drugs is difficult because one needs to specify how the rate of resistance evolution to each drug is affected by the number of drugs in use (something for which virtually no data are available). Under some assumptions, however, allowing for simultaneous multiple drug use results in a model that is formally identical to the model presented here (see Appendix). This shows that, under some conditions, all of the insights obtained in our simplified model carry over to this more complex situation. It also strongly suggests that, even for other simultaneous drug use scenarios, where an exact correspondence with the simple model no longer holds, the important processes elucidated here will continue to operate. Other, additional, processes might then occur as well, however, and this is therefore an important area for future research.

Our analysis does not consider how investments can be delivered to the system. Additive and multiplicative changes were used to determine the importance of speeding up drug arrivals relative to slowing evolution of drug resistance, which implies that changing either component in the same manner requires the same amount of resources or effort. However, it might be less costly or easier to change drug lifespan than to change drug development or vice versa. Further research in this area might consider assessing strategies using empirically supported cost functions for investments [4, 20-22].

Even though we provide evidence that slowing evolution of drug resistance is more beneficial to the effective management of resistance than speeding up drug development, we stress that this in no way negates the importance of pharmaceutical innovation and drug discovery. Indeed, the two approaches to resistance management need not always be traded off against one another. For example, slowing the evolution of drug resistance is, in many ways, fundamentally linked to drug development. Investment and advancement in drug development might lead to smarter drugs with enhanced efficacy and greater longevity. Thus our intention is not to suggest that there is, by necessity, an antagonism between the two approaches. The results do demonstrate, however, that a greater emphasis on evolution management might yield promising results for this pressing problem.

**Acknowledgements:** We are grateful to Andrew Read for helpful comments and critical reading of the manuscript. This research was supported by a NSERC Graduate Scholarship (N.S.M.), a Discovery Grant (T.D.) and the Canada Research Chairs (T.D.).

**Figures and Figure Legends:**

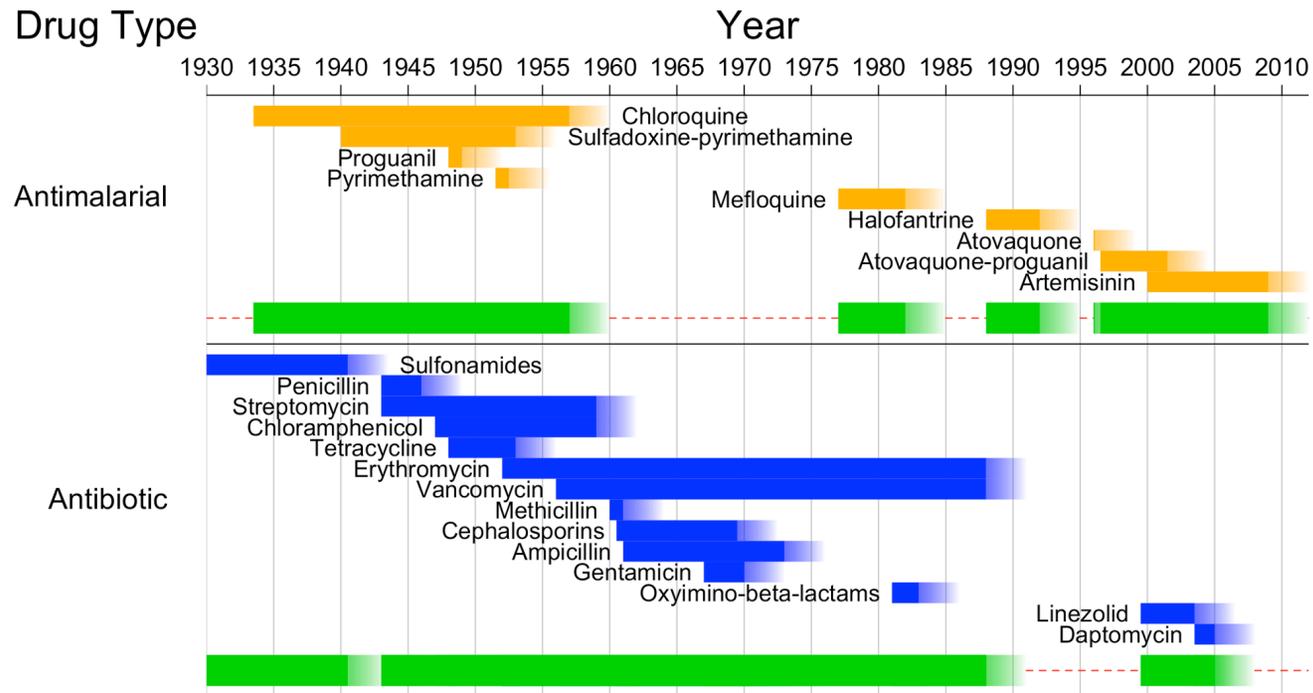

**Figure 1.** Drug supply timeline for antimalarials and anitbiotics. This gives the time of drug introduction and the subsequent evolution of resistance (6-12). We show the observed periods of time that effective treatment is available (green bars) separated by periods during which there is a risk of ineffective treatment (dashed red-lines) as a result of resistance. The color fading is meant to show that first observation of resistance does not absolutely equate with complete loss of treatment efficacy.

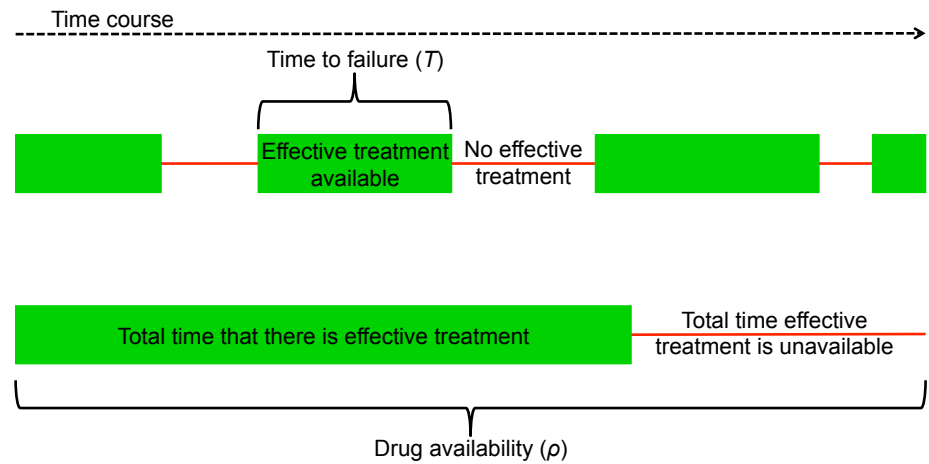

**Figure 2.** Illustration of drug supply. A schematic timeline defining the time to failure, $T$, and drug availability, $\rho$. 'Effective treatment available' means that there is at least one effective drug in the portfolio.

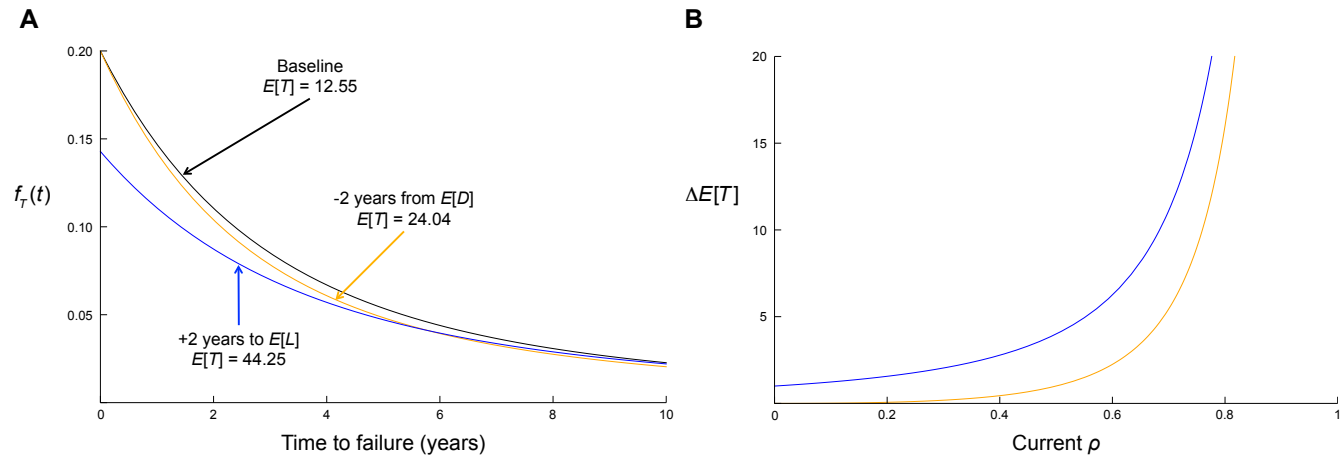

**Figure 3.** Effects on time to failure from enhancing drug development and slowing evolution. (A) The effect on the time to failure density $f_T(t)$ when the average time between drug arrivals is reduced by 2 years (orange line) and the average time to evolve resistance is extended by 2 years (blue line) compared to baseline conditions (black line; $E[L] = 5$ years, $E[D] = 8.3$ years). (B) The change in expected time to failure ($E[T]$) resulting from additive perturbations in $E[L]$ (blue line) and $E[D]$ (orange line) plotted for varying current drug availability $(\rho)$.

**Appendix Contents**

# 1 – Data

Data were collected from various sources to create a timeline of drug supply for antimalarials and antibiotics (Table A1). The data consist of the date of introduction, and the date of first recorded resistance. When there existed multiple estimates for the same drug from a single source, we report the earliest date that the drug was introduced and the earliest date that resistance was first observed (this occurred when there was region-specific data on drug introduction or resistance). While it is reported that resistance has developed to every antimalarial (with the possible exception of artemisinin therapies [3]), we excluded data on Lapdap, Amodiaquine, and Primaquine because the determination of drug resistance for these antimalarials was inconclusive (e.g., due to a lack of evidence or because of confounding factors in treatment failure such as cross resistance, side-effects, compliance and drug withdrawal [23-25]).

We also used the available antimalarial data to give an example parameterization of the model. We stress, however, that this is intended merely as an illustrative case. All of the results reported in the main text are general, and independent of the disease in question as well as the specific parameter values used. We nevertheless chose to include an example parameterization based on data because it helps to clarify the relevance of the general results and to make them more concrete.

Coming up with a suitable parameterization from the data is difficult because our simplified model lacks some of the features of the real data. Specifically, in the data many drugs have overlapping lifespans (elapsed time from date of drug introduction to date of observation of resistance). This suggests that more than one drug is used at a time during these periods. Our simple model assumes that one drug is used at a time (although Section 4 generalizes these results) and therefore we need to extract suitable estimates from the data under this assumption. For example, we cannot use the observed drug lifespans since the sum of individual drug lifespans would be greater than the observed time to failure, misrepresenting the current state of drug supply. As a result we employed two different approaches (see below) to estimate the time to evolve resistance from the data. The distribution of time to evolve resistance is almost identical under both approaches, and resembles an exponential distribution with rate parameter $\beta = 0.2$ (see Figure A1). The distribution of time between drug arrivals also resembles an exponential distribution with rate parameter $\alpha = 0.12$, supporting an assumption based on results from [17] (see Figure A1).

### *Approach 1: Dividing overlapping periods by the number of drugs*

We can estimate the time to resistance of individual drugs by dividing overlapping periods by the number of drugs and summing the times that a particular drug was used. Section 4 describes how this approach can be used to generalize the one-drug-at-a-time model to a multiple drug scenario. In effect, this gives an estimate of the time to evolve resistance for each drug in the absence of any other drug.

*Approach 2: Time to evolve resistance from earliest drug arrival to first drug failure*

Alternatively, we can consider time to evolve resistance as the time from the earliest drug arrival to the time that resistance is first observed to any drug in an overlapping period. Using this approach, estimates of time to resistance are not necessarily for a specific drug, but instead reflect the time to resistance for a group of drugs (and the factors experienced by these drugs) that are in use during an overlapping period.

*Simulating the current and future state of antimalarial supply*

Antimalarial drug supply was simulated using the parameter estimates from the antimalarial data as described above. Figure A2 shows how periods of time to failure and drug availability observed from 1930-2012 in the antimalarial data agree with a characteristic simulation of the current state of drug supply. We can also envision what might happen to antimalarial supply in the future if we slowed resistance or enhanced drug development. In this case, we chose to compare an increase of 2 years in the mean time to evolve resistance with a decrease of 2 years in the mean time between drug arrivals. Figure A3 shows that increasing the mean time to evolve resistance results in significantly longer periods of time to failure, which also means that drugs will be available for a longer fraction of time over the next 200 years.

Again we stress, however, that these simulation results are merely meant to be an illustrative example. The findings from the model are independent of the details of this simulation, and apply for any distribution of time to resistance, regardless of its shape or parameter values. The model also allows for the use of any (consistent) measure of resistance or drug failure since time to resistance is treated as a random process. In this example, the data report the first observation of resistance but we might wish to measure drug lifespan as the time until resistance is observed at some threshold frequency instead.

## 2 – Analysis

All models presented in the text are analogous to models from queueing theory. There are abundant results available for analyzing their behavior [26]. In what follows we make use of results found in [26].

*Exponentially Distributed Time to Evolve Resistance*

When the time to evolve resistance is exponentially distributed, the drug supply model is analogous to an M/M/1 queue [26]. Using $f_T(t)$ to denote the probability density of the time to failure, and $\tilde{f}_T(z)$ for its *Laplace-Stieltjes transform* (LST), standard results [26] demonstrate that

$$\tilde{f}_T(z) = \frac{\alpha + \beta + z - \sqrt{(\alpha + \beta + z)^2 - 4\beta\alpha}}{2\alpha} \qquad \text{(Equation 1)}$$

where $E[D] = 1/\alpha$ and $E[L] = 1/\beta$ (Equation 1 is also derived below in the case of an arbitrary distribution for the time to evolve resistance).
Inverting the transform gives the density [26]

$$f_T(t) = \sqrt{\frac{\beta}{\alpha}} \frac{e^{-(\alpha+\beta)t} I_1(2\sqrt{\alpha\beta}t)}{t} \tag{Equation 2}$$

where $I_1(x)$ is a modified Bessel function of the first kind with order 1. Specifically, $I_1(x)$ can be defined by the integral formula

$$I_1(x) = \frac{1}{\pi} \int_0^\pi e^{x\cos\theta} \cos\theta \, d\theta. \tag{Equation 3}$$

Drug availability, $\rho$, is the long run proportion of time an effective drug is available. In the context of queueing theory this is referred to as the "load" or "traffic intensity". Standard results [26] show that

$$\rho = E[L]/E[D]. \tag{Equation 4}$$

Equation 4 can be understood by recognizing that drug availability is the long-term time spent with at least one drug (which is, on average, $E[T]$) divided by the length of the cycle (which is, on average, $E[T] + E[D]$). In other words, the ratio

$$\frac{E[T]}{E[T] + E[D]}.$$

Using Equation 2 we can calculate $E[T] = E[L]/(1 - E[L]/E[D])$. Substituting this into the above ratio then gives Equation 4.

We can now determine the effect of each intervention on Equation 2 and Equation 4. The density (Equation 2) is a decreasing function of $t$ and changes in $E[D] = 1/\alpha$ or $E[L] = 1/\beta$ affect the density differently. For example,

$$\lim_{t \to 0} f_T(t) = \frac{1}{E[L]},$$

shows that increasing the expected time to evolve resistance (either additively or multiplicatively) decrease the probability density at $t = 0$ whereas decreasing the expected time between drug arrivals has no effect at this point. Consequently, given that the density function is continuous, interventions that target evolution will shift more probability density from near zero to larger values of $T$ than will interventions that target drug development.

For the drug availability, Equation 4, we consider additive and multiplicative changes in turn. An additive change of size $x$ will either add $x$ to the mean time to evolve resistance or subtract $x$ from the mean time between drug arrivals. A multiplicative change of size $y>1$ will either multiply the mean time to evolve resistance by $y$ or divide the mean time between drug arrivals by $y$.

For additive changes of size $x$ the inequality

$$\frac{E[L]+x}{E[D]} > \frac{E[L]}{E[D]-x}$$

holds for all $x$ that satisfy the assumption that $\rho$ is less than 1. Alternatively, multiplicative changes result in equal benefits to drug availability as

$$\frac{E[L] \times y}{E[D]} = \frac{E[L]}{E[D]/y}.$$

Therefore, when evolution of drug resistance is slowed, effective treatment will be available for a fraction of time greater than or equal to the effect from enhancing drug development.

## *Arbitrary Distribution for the Time to Evolve Resistance*

When time to evolve resistance has an arbitrary distribution, the drug supply model is analogous to an M/G/1 queue [26]. We can derive an equation for the *Laplace-Stieltjes transform* (LST) of the time to failure distribution as follows.

Any time to failure period begins with the arrival of a new drug into the portfolio. While this drug is in use, additional new drugs might be added to the portfolio. Suppose, for example, there are $K$ additional drugs added during the time that the first drug is being used ($K$ is a random variable). Each of these additional drugs will eventually be used, and each will, themselves, spawn a time to failure period that has the same distribution as that of the first drug. In this way we can write

$$T = L + T_1 + \cdots + T_K$$

where $L$ is the time to evolve resistance for the first drug, and the $T_i$'s are the time to failure periods for the additional $K$ drugs that arrive before the first drug is abandoned. Using $f_L(t)$ and $f_T(t)$ to denote the probability densities for the time to evolve resistance and time to failure respectively, the LST of $f_T(t)$ is $\tilde{f}_T(z) = E[e^{-zT}]$. Conditioning on the lifespan of the first drug we obtain

$$\tilde{f}_T(z) = \int_0^\infty f_L(t) E[e^{-zT} | L = t] dt,$$

and further conditioning on the number of new drug arrivals, $K$, during this time gives

$$\tilde{f}_T(z) = \int_0^\infty f_L(t) \left( \sum_{k=0}^\infty E[e^{-z(L+T_1+\cdots+T_K)} \mid L=t, K=k] P(K=k \mid L=t) \right) dt$$

$$= \int_0^\infty f_L(t) e^{-zt} \left( \sum_{k=0}^\infty E[e^{-z(T_1+\cdots+T_k)}] P(K=k \mid L=t) \right) dt$$

$$= \int_0^\infty f_L(t) e^{-zt} \left( \sum_{k=0}^\infty E[e^{-z(T_1+\cdots+T_k)}] \frac{(\alpha t)^k e^{-\alpha t}}{k!} \right) dt$$

where the final equality makes use of the fact that $K$ is Poisson distributed. Now, using the fact that the $T_i$'s are iid, we obtain

$$\tilde{f}_T(z) = \int_0^\infty f_L(t) e^{-zt} \left( \sum_{k=0}^\infty E[e^{-zT}]^k \frac{(\alpha t)^k e^{-\alpha t}}{k!} \right) dt$$

$$= \int_0^\infty f_L(t) e^{-zt} \left( \sum_{k=0}^\infty \tilde{f}_T(z)^k \frac{(\alpha t)^k e^{-\alpha t}}{k!} \right) dt.$$

Finally, noting that $\sum_{k=0}^\infty z^k \frac{(\alpha t)^k e^{-\alpha t}}{k!} = e^{(z-1)\alpha t}$, this last expression simplifies to

$$\tilde{f}_T(z) = \int_0^\infty f_L(t) e^{-zt} e^{(\tilde{f}_T(z)-1)\alpha t} dt$$

$$= \int_0^\infty f_L(t) e^{-(z+\alpha - \alpha \tilde{f}_T(z))t} dt$$

or

$$\tilde{f}_T(z) = \tilde{f}_L\left(z + \alpha - \alpha \tilde{f}_T(z)\right) \qquad \text{(Equation 5)}$$

Equation 5 is Takács functional equation relating the LST of the distribution of time to failure, $T$, to the LST of the distribution of time to evolve resistance, $L$.

We can now calculate any moment of interest for the distribution of the time to failure, in terms of the moments of the distribution of time to evolve resistance. In particular, $E[T] = -\tilde{f}_T'(0)$, and therefore

$$E[T] = \frac{E[L]}{1 - E[L]/E[D]}. \qquad \text{(Equation 6)}$$

Notice that this is identical to the expression for E[$T$] obtained in the case of exponentially distributed time to evolve resistance, and therefore Equation 4 for drug availability remains valid even when the time to evolve resistance has an arbitrary distribution.

Incidentally, we can also derive Equation 1 from Equation 5 by using the fact that, when $L$ is exponentially distributed with parameter $\beta$, the LST of $L$ is $\tilde{f}_L(z) = \beta/(\beta+z)$. Substituting this into Equation 5 gives

$$\tilde{f}_T(z) = \frac{\beta}{\beta + z + \alpha - \alpha \tilde{f}_T(z)}.$$

Solving this quadratic equation for $\tilde{f}_T(z)$ gives Equation 1.

When the distribution of $L$ is arbitrary, we can consider the effect that additive and multiplicative changes have on any moment of the time to failure distribution. For the purposes of this analysis, we have chosen to focus on the first moment, the expected time to failure, $E[T]$ (Equation 6). The expected time to failure can be broken up into two parts: the average number of drugs used before failure, $E[N] = (1 - E[L]/E[D])^{-1}$ and the average lifespan of each drug, $E[L]$, such that $E[T] = E[N] \times E[L]$. The likelihood that another drug arrives before failure will increase as a result of slowing the time to evolve resistance or decreasing the time between drug arrivals. An additive change of size $x$ increases the average number of drugs used before failure and, in particular, the inequality

$$\left(1 - \frac{E[L] + x}{E[D]}\right)^{-1} > \left(1 - \frac{E[L]}{E[D] - x}\right)^{-1}$$

holds for all $x$ that satisfy the assumption that $\rho$ is less than 1. Alternatively, there is an equivalent increase in the average number of drugs used before failure when evolution is slowed and drug development is enhanced multiplicatively since

$$\left(1 - \frac{E[L] \times y}{E[D]}\right)^{-1} = \left(1 - \frac{E[L]}{E[D]/y}\right)^{-1}.$$

This shows that increasing the mean time to evolve resistance will result in at least as many drugs used before failure as decreasing the mean time between drug arrivals. In addition, since increasing the mean time to evolve resistance also increases average drug lifespan, the total effect on average time to failure is that much greater when resistance is slowed than when drug development is enhanced.

We also show how increasing the mean time to evolve resistance or decreasing the mean time between drug arrivals affects the whole time to failure distribution. The drug supply system was simulated for a constant rate of drug arrivals by randomly selecting the time between drug arrivals from an exponential distribution with rate $\alpha$. The time it takes to evolve resistance to a drug was randomly selected from a normal distribution with expected time to evolve drug resistance, $E[L] = \mu$, and standard deviation $\sigma$. Figure A4 shows that this entire distribution shifts more toward longer durations of time to failure when the evolution of drug resistance is slowed than when the rate of drug arrivals is sped up, thereby producing the larger change in expected value of *T* described above.

## 3 – Variable Rate of Drug Development

*Model*

We will also consider a variable rate of drug arrivals so that drugs arrive at a normal rate, $\alpha_D$, when there are lots of drugs in the drug portfolio and a fast rate, $\alpha_F$, when there are few effective drugs available. The normal rate of drug arrivals was varied to determine the effect of speeding up drug development. As before, these results were compared against the effect of slowing evolution of resistance. In these simulations, the proportion of time that drugs arrived at their respective rates was also considered.

The available empirical evidence suggests that annual drug production has been constant for over 55 years, remaining unchanged in spite of increased investment [17]. Even so, by including a fast rate, the time between drug arrivals is not necessarily longer than the time it takes resistance to evolve, relaxing the $E[D]>E[L]$ assumption of the previous model. Results from this simulations also show that slowing the evolution of resistance has a greater impact on time to failure and drug availability than does speeding up the normal rate of drug arrivals.

*Results*

Numerical results show that slowing the evolution of resistance benefits time to failure and drug availability more than speeding up the normal rate of drug arrivals (Table A2). Slowing evolution of drug resistance also decreased the proportion of time that drugs arrived according to a fast rate of development (Figure A5). This suggests that drug arrivals were occurring close to a maximum rate, so that any further increase in the normal rate of drug arrivals would have little effect on system performance. In contrast, when evolution was slowed, a normal rate of drug arrival was used for a greater proportion of time implying that the average rate of drug arrivals slowed down when $E[L]$ increased.

## 4 – Generalizing to Multiple Drugs

Thus far, we have presented a model of drug supply that assumes one drug or drug therapy is used at a time. While this assumption is supported by current treatment guidelines and practice for malaria (wherein a single first line therapy is recommended for treatment and replaced when resistance emerges [13, 27]) in general, it is likely that more than one drug will be used to treat a disease when available. Note, by this we mean that individuals are still given a single drug, but different individuals might be given different drugs simultaneously.

To begin it is helpful to first consider how simultaneous drug use affects the total time required to evolve drug resistance for a set of drugs, relative to that required if the drugs in this set were used one at a time (i.e., sequentially). There are three possibilities: (i) conservation of evolution, (ii) compression of evolution, and (iii) expansion of evolution. These are defined as follows.

Suppose there are $n$ drugs currently available. We say that there is conservation of evolution if the total time it takes for resistance to evolve to all $n$ drugs when used simultaneously is the same as the time required for resistance to evolve to all $n$ drugs when used sequentially. Alternatively, we say that there is compression of evolution if the time taken under simultaneous drug use is shorter, and expansion of evolution if it is longer. We will show that this provides a heuristic framework with which to examine multiple drug use.

It can be shown, using results from work-conserving queueing theory [28], that a model with simultaneous drug use is formally equivalent to a one-drug-at-a-time model when there is conservation of evolution. We propose a mechanistic basis by which this occurs. Suppose that the rate of evolution for individual drugs depends on the number of drugs in use, $n$, such that the rate of evolution for each drug is slowed by a factor $1/n$ relative to the rate of resistance evolution if it were the only drug in use. Then it can be shown that the distribution of time to failure is unchanged from a one drug at a time model, as this is analogous to a work-conserving queue [28]. In a similar way, we can account for drugs used at different frequencies by assuming that the rate of evolution of resistance to each is slowed by a factor equal to its frequency of use.

These ideas reveal that, under certain conditions (namely, conservation of evolution) the results of the simple model of the text are identical to those for a model that allows simultaneous multiple drug use. Therefore, increasing the mean time to evolve resistance is more effective than decreasing the mean time between drug arrivals in an evolution-conserving scenario of multiple drug use. This also provides a natural point of departure for exploring cases in which evolution is not conservative.

When multiple drug use results in compression of evolution, this means that there are interactions among drugs that result in a reduction of the time required to evolve resistance. This effect might arise from cross-resistance, wherein the mechanism of resistance to one drug also confers resistance to another drug, effectively reducing its lifespan. This means that it is more important to slow resistance than to enhance drug development, as simply using more drugs will reduce the lifespan of each drug.

Alternatively, when multiple drug use results in the expansion of evolution, this means that there are interactions among drugs that bring about an increase in the time required for resistance evolution. In large part, it is precisely these interactions that have succeeded in the development of combination therapies as a single treatment for individuals. At the population level, drug mixing has been suggested as a way of spatially varying the drug environment so that it is more difficult for microbes to evolve resistance to any one drug [29].

# 5 – Supplementary Tables

**Table A1.** Data for antimalarials and antibiotics. The values indicate the time of drug introduction and first observation of drug resistance. The reference for each drug is provided; all data is from published sources.

| Drug | Drug Type | Introduced | Resistance | Source |
|---|---|---|---|---|
| Chloroquine | Antimalarial | 1933.5 | 1957 | (a) |
| Sulfadoxine-pyrimethamine | Antimalarial | 1940 | 1953 | (a) |
| Proguanil | Antimalarial | 1948 | 1949 | (b) |
| Pyrimethamine | Antimalarial | 1951.5 | 1952.5 | (c) |
| Mefloquine | Antimalarial | 1977 | 1982 | (d) |
| Halofantrine | Antimalarial | 1988 | 1992 | (b) |
| Atovaquone | Antimalarial | 1996 | 1996 | (d) |
| Atovaquone-proguanil | Antimalarial | 1996.5 | 2001.5 | (c) |
| Artemisinin | Antimalarial | 2000 | 2009 | (a) |
| Sulfonamides | Antibiotic | 1930 | 1940.5 | (e) |
| Penicillin | Antibiotic | 1943 | 1946 | (f) |
| Streptomycin | Antibiotic | 1943 | 1959 | (f) |
| Chloramphenicol | Antibiotic | 1947 | 1959 | (f) |
| Tetracycline | Antibiotic | 1948 | 1953 | (f) |
| Erythromycin | Antibiotic | 1952 | 1988 | (f) |
| Vancomycin | Antibiotic | 1956 | 1988 | (f) |
| Methicillin | Antibiotic | 1960 | 1961 | (f) |
| Cephalosporins | Antibiotic | 1960.5 | 1969.5 | (e) |
| Ampicillin | Antibiotic | 1961 | 1973 | (f) |
| Gentamicin | Antibiotic | 1967 | 1970 | (g) |
| Oxyimino-beta-lactams | Antibiotic | 1981 | 1983 | (g) |
| Linezolid | Antibiotic | 1999.5 | 2003.5 | (e) |
| Daptomycin | Antibiotic | 2003.5 | 2005 | (e) |

(a) Shetty P (2012) Malaria – the numbers game. *Nature* 484:A14-A15.
(b) Thayer AM (2005) Fighting malaria. *Chemical and Engineering News* 83:69-82.
(c) Hyde JE (2005) Drug-resistant malaria. *Trends Parasit.* 21:494-498.
(d) Wongsrichanalai C, Pickard AL, Wernsdorfer WH, Meshnick SR (2002) Epidemiology of drug-resistant malaria. *Lancet Infect Dis* 2:209-218.
(e) Clatworthy AE, Pierson E, Hung DT (2007) Targeting virulence: a new paradigm for antimicrobial therapy. *Nat Chem Biol* 3:541-548.
(f) Palumbi SR (2001) Humans as the world's greatest evolutionary force. *Science* 293:1786-1790.
(g) Jacoby GA (2009) in *Antimicrobial Drug Resistance*, ed Mayers DL (Humana Press, New York), pp 3-7.

**Table A2.** Performance measures for a variable rate of drug development. Results from simulations (n = 100) with a variable rate of drug arrivals over a time interval of 300 years. The mean time between drug arrivals was 5 years ($E[F] = 5$) when there were less than 3 drugs in the system at the time of the last drug arrival. An additive change affects the mean time to evolve resistance or mean normal time between drug arrivals by 2 years, while a multiplicative change doubles the mean time to resistance or halves the mean normal time between drug arrivals. Average drug availability is the drug availability realized for each simulation averaged over all simulations. Average time to failure and down time is the average time to failure and down time for each simulation averaged over all simulations. Average proportion of time that drugs arrived from a particular rate is equal to the proportion of time the rate is used for each simulation averaged over all simulations.

|  | Initial conditions | Additive change | | Multiplicative change | |
|---|---|---|---|---|---|
| *(Mean time between drug arrivals, Mean time to evolve resistance)* | 10, 3 | 10, 5 | 8, 3 | 10, 6 | 5, 3 |
| **Average drug availability** | 0.523 | 0.737 | 0.541 | 0.813 | 0.592 |
| **Average time to failure (years)** | 6.227 | 18.629 | 6.444 | 29.652 | 7.764 |
| **Average down time (years)** | 5.721 | 6.814 | 5.479 | 7.024 | 5.165 |
| **Average proportion of time fast rate used** | 0.746 | 0.471 | 0.741 | 0.358 | 0.731 |

# 6 – Supplementary Figures

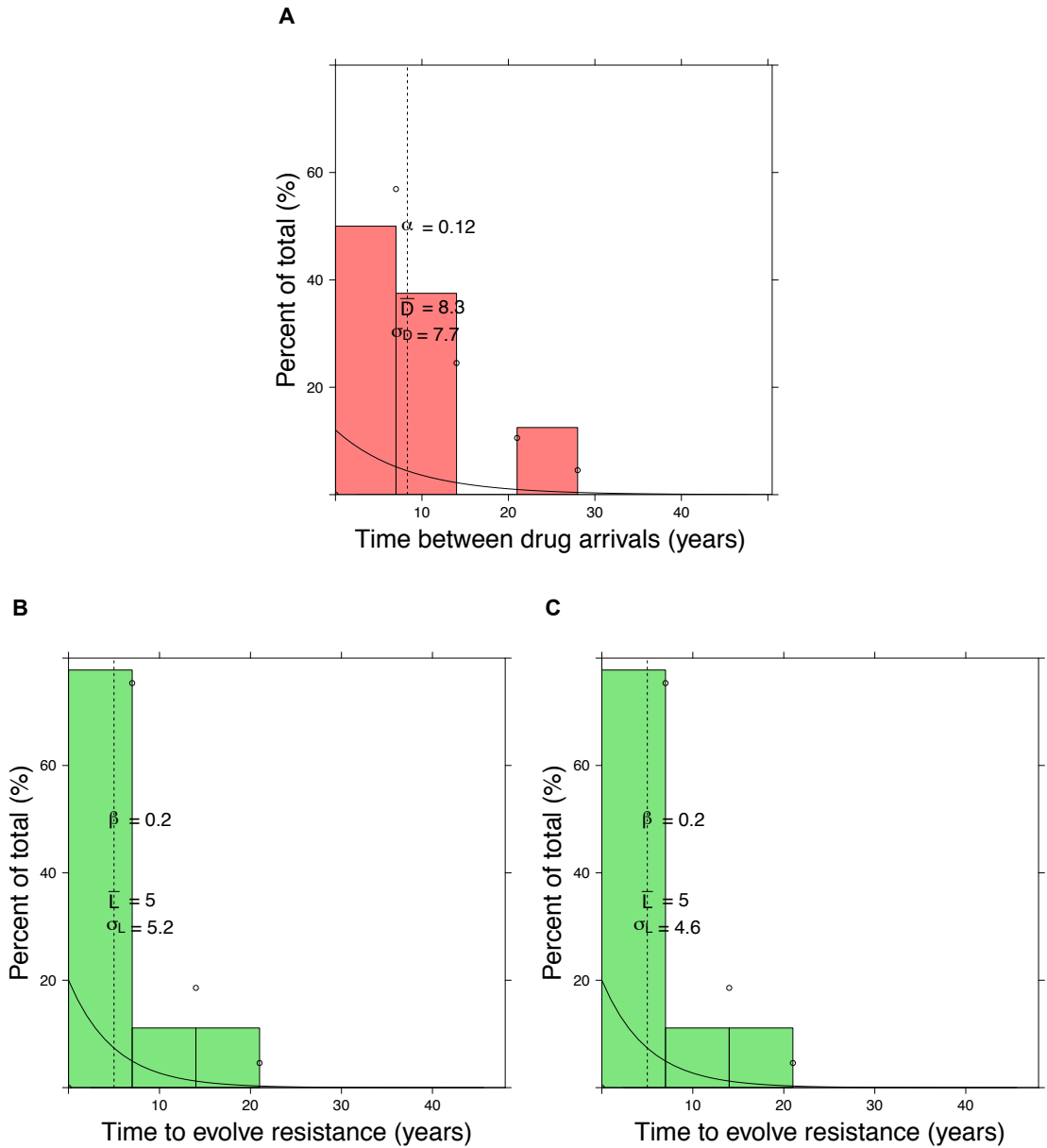

**Figure A1.** Histograms of drug supply parameters from antimalarial data. (A) The distribution of time between drug arrivals ($\alpha = 0.12$, $\bar{D} = 8.3$ years, $\sigma_D = 7.7$ years). The distribution of time to evolve resistance was estimated via two approaches: (B) by dividing overlapping regions by the number of drugs ($\beta = 0.2$, $\bar{L} = 5$ years, $\sigma_L = 5.2$ years) or (C) using the time from earliest drug arrival to first drug failure ($\beta = 0.2$, $\bar{L} = 5$ years, $\sigma_L = 4.6$ years). Using the estimate of the rate parameter in each panel, the solid line is the exponential density plot and the open circles give the probability (%) within the interval defined by the width of each bar. The dashed line is the average value of the variable of interest.

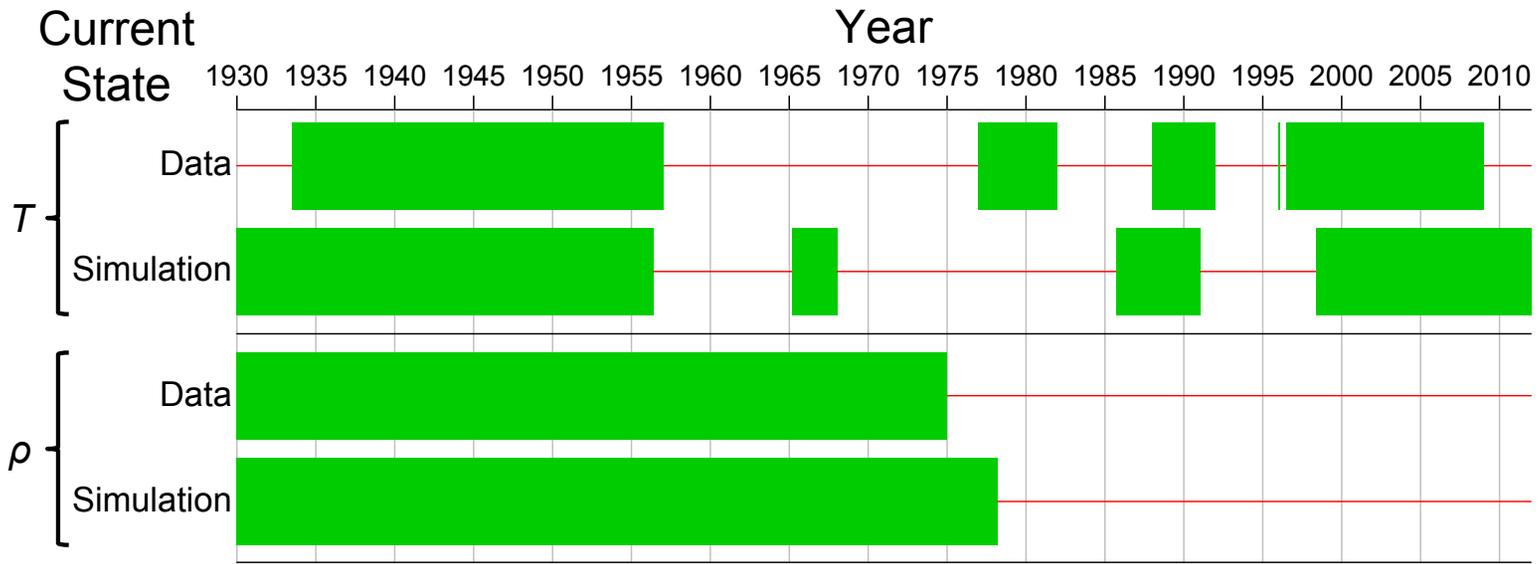

**Figure A2.** The current state of antimalarial supply. This shows the time to failure ($T$; green bars) and drug availability ($\rho$) observed from data and from simulation using parameter estimates ($E[L] = 5$, $E[D] = 8.3$).

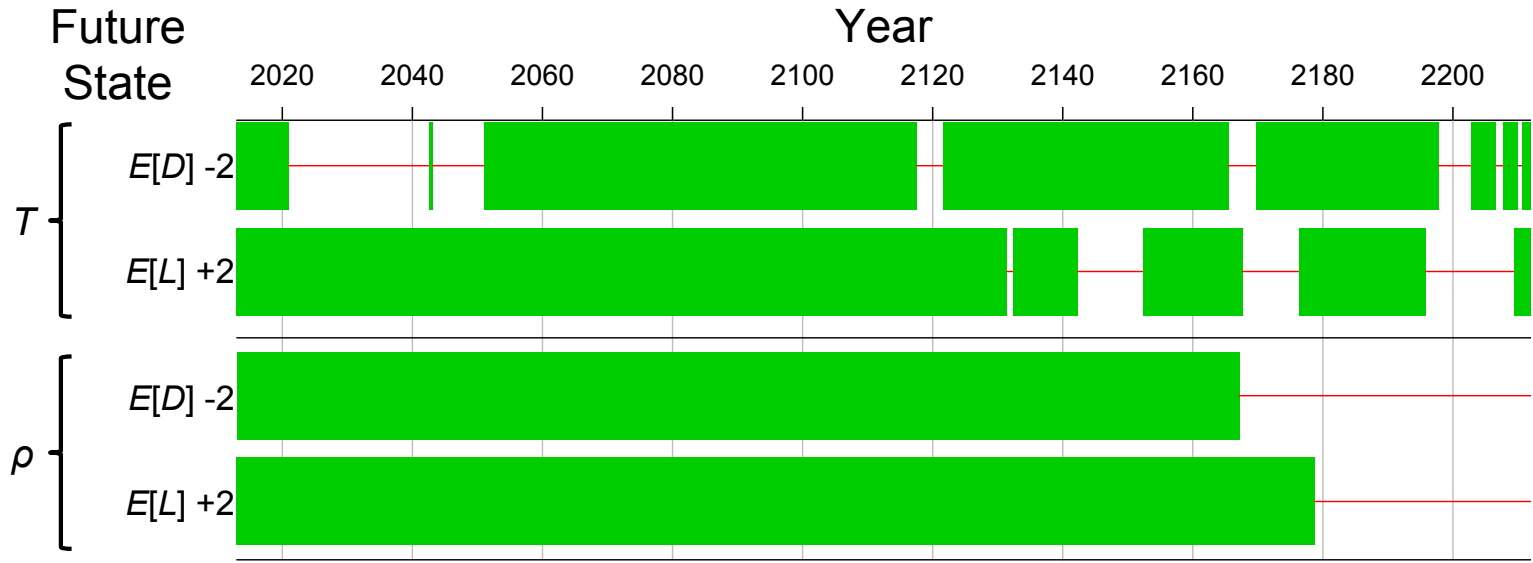

**Figure A3.** Characteristic realizations of future antimalarial supply. A simulation over the next 200 years when the mean time between drug arrivals is reduced by 2 years or the mean time to evolve resistance is increased by 2 years (compared to initial conditions: $E[L] = 5$, $E[D] = 8.3$).

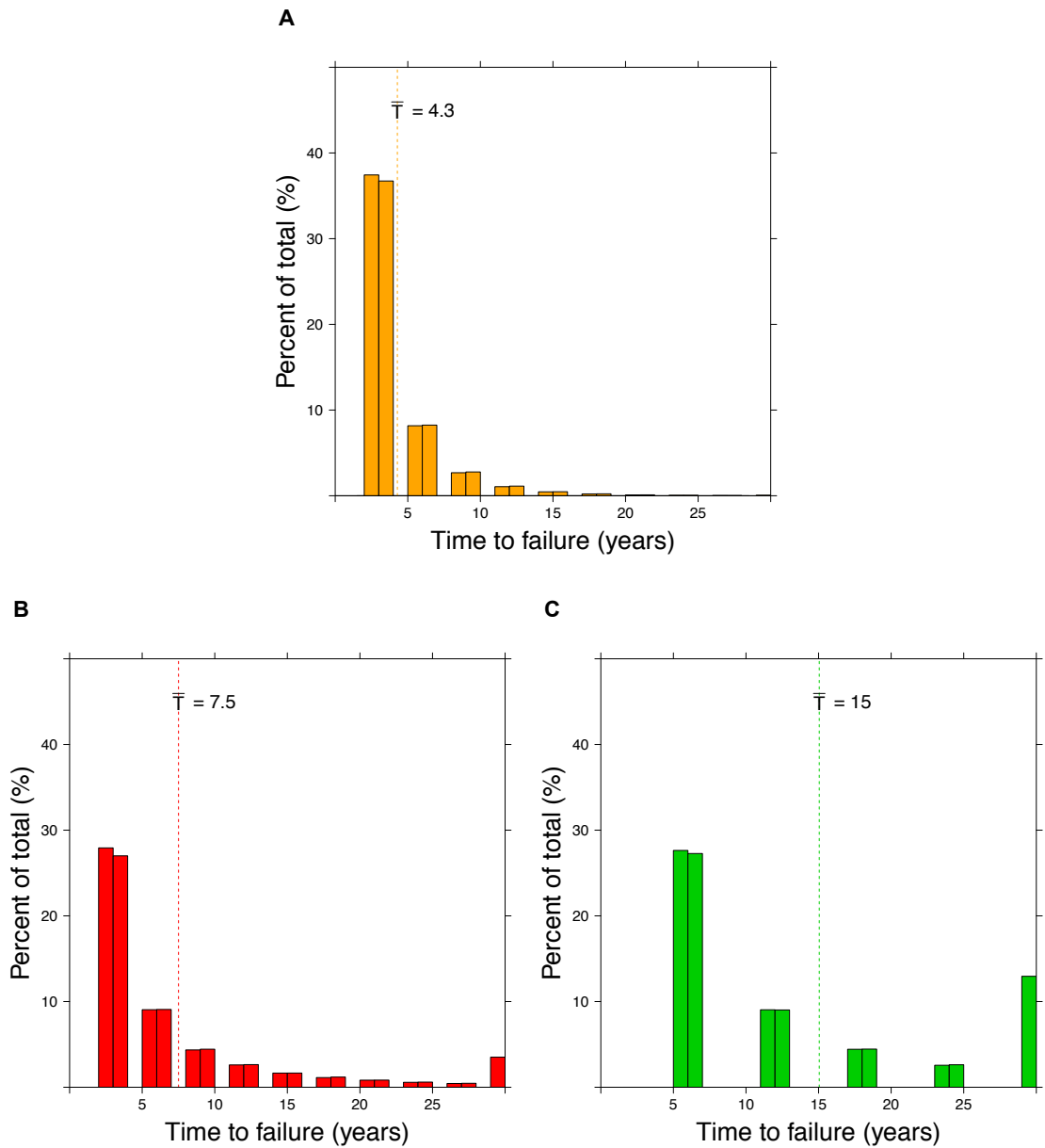

**Figure A4.** Histograms of time to failure from simulation. (A) For initial conditions ($E[D] = 10$, $E[L] = 3$, $\sigma_L = 0.1$), and after a multiplicative change is applied: (B) halving the mean time between drug arrivals ($E[D] = 5$, $E[L] = 3$, $\sigma_L = 0.1$) or (C) doubling the mean time to evolve drug resistance ($E[D] = 10$, $E[L] = 6$, $\sigma_L = 0.1$). The last bar in each histogram shows lengths of time to failure greater than or equal to 29 years and the dashed line indicates the average length of time to failure in each panel.

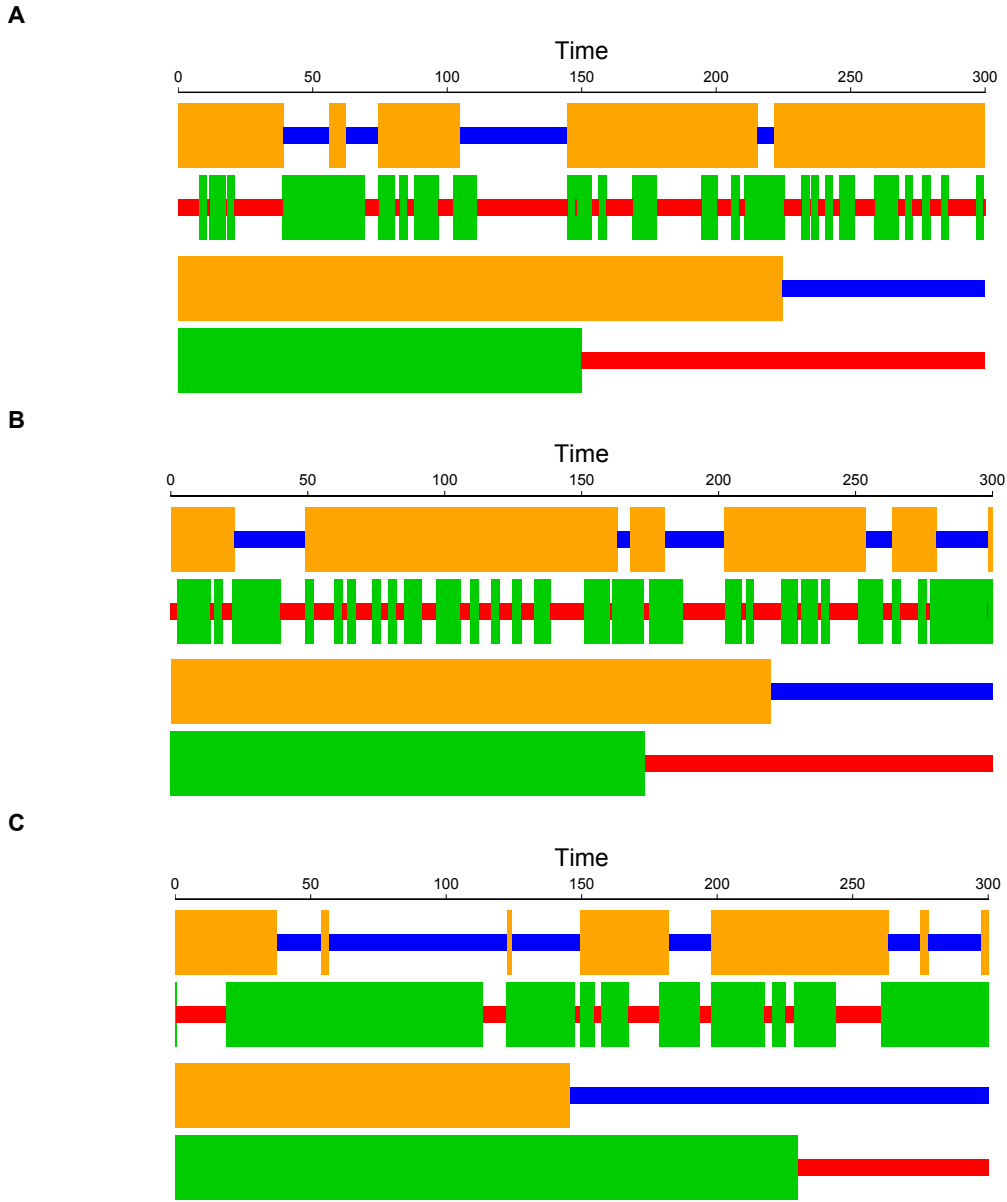

**Figure A5.** Characteristic simulations of a variable rate of drug development. The pattern of time to failure (green bars) and down time (red lines) along with the periods that a fast rate (orange bars) and normal rate (blue lines) of drug arrivals were used over a time interval for (A) initial conditions ($E[D] = 10$, $E[F] = 5$, $E[L] = 3$, $\sigma_L = 0.1$), and after an additive change is applied: (B) subtracting 2 from the mean normal time between drug arrivals ($E[D] = 8$, $E[F] = 5$, $E[L] = 3$, $\sigma_L = 0.1$) or (C) adding 2 to the mean time to evolve drug resistance ($E[D] = 10$, $E[F] = 5$, $E[L] = 5$, $\sigma_L = 0.1$). The mean time between drug arrivals when there are less than 3 drugs in the drug portfolio is held constant at 5 ($E[F] = 5$). Drug availability is shown at the bottom of each plot as the sum of time to failure (green area) relative to the sum of down time (red area) over the time interval. The proportion of time that drugs arrived from a fast rate (orange area) rate relative to a normal rate (blue area) is also shown.